\documentclass[journal,comsoc]{IEEEtran}
\usepackage[T1]{fontenc}

\usepackage{amssymb,amsmath,amsfonts,amsthm}
\usepackage{mathrsfs}			
\usepackage{cite}
\usepackage{array}
\usepackage{commath}
\usepackage{cite}
\usepackage{tabularx}
\usepackage[table]{xcolor}
\usepackage{mathtools}
\usepackage{mathtools}
\usepackage{subfigure}
\usepackage{mathtools}
\usepackage{graphicx}
\usepackage{bbm}
\usepackage[utf8]{inputenc}
\usepackage{algorithm}

\newtheorem{theorem}{Theorem}
\newtheorem{lemma}{Lemma}

\newcommand{\maximize}{\mathop{\rm maximize}\limits}

\IEEEoverridecommandlockouts
\ifCLASSINFOpdf
\else
\fi
\usepackage[cmintegrals]{newtxmath}
\begin{document}
	\title{Dynamic Resource Allocation in Co-Located and Cell-Free Massive MIMO}
	
	\author{Zheng~Chen,~\IEEEmembership{Member,~IEEE}, Emil~Bj\"{o}rnson,~\IEEEmembership{Senior Member,~IEEE}, and~Erik~G.~Larsson,~\IEEEmembership{Fellow,~IEEE}
		\thanks{Z. Chen, E. Bj\"{o}rnson and E. G. Larsson are with the Department of Electrical Engineering (ISY), Link\"{o}ping University, Link\"{o}ping, Sweden (email: zheng.chen@liu.se, emil.bjornson@liu.se, erik.g.larsson@liu.se). A part of this work was presented at the IEEE International Conference on Communications (ICC) 2019. This work was supported in part by ELLIIT, CENIIT, and the Swedish Foundation for Strategic Research.}}
	\maketitle

\begin{abstract}	
In this paper, we study  joint power control and scheduling in
uplink massive multiple-input multiple-output (MIMO) systems with
randomly arriving data traffic. We consider both co-located and
Cell-Free (CF) Massive MIMO, where the difference lies in whether the
antennas are co-located at the base station or spread over a wide
network area. The data is generated at each user according to an
individual stochastic process. Using Lyapunov optimization techniques,
we develop a dynamic scheduling algorithm (DSA), which decides at each
time slot the amount of data to admit to the transmission queues and
the transmission rates over the wireless channel. The proposed
algorithm optimizes the long-term user throughput under various
fairness policies while keeping the transmission queues stable.
Simulation results show that the state-of-the-art power control
schemes developed for Massive MIMO with infinite backlogs can fail to
stabilize the system even when the data arrival rates are within the
network capacity region. Our proposed DSA shows advantage in providing
finite delay with performance optimization whenever the network can be
stabilized.
\end{abstract}

\section{Introduction}

Massive multiple-input multiple-output (MIMO) is one of the key
technologies in 5G \cite{mmimo_mag,Parkvall2017a}. By deploying base
stations (BSs) equipped with many antennas, spatial multiplexing can
be utilized to serve a large number of users on the same
time-frequency resource. The rates and energy efficiency of the
network can be significantly improved by Massive MIMO compared to in
traditional MIMO systems \cite{marzetta2016fundamentals,
  massivemimobook}. In a co-located Massive MIMO network, all 
antennas are located at one single base station in each
cell. Cell-edge users usually suffer from much higher path loss than
cell-center users. Recently, Cell-Free (CF) Massive MIMO has emerged
as an alternative implementation of the Massive MIMO concept, with the potential
to provide better coverage probability. In such systems, the access
points (APs) with either one or multiple antennas are distributed at
different places and  jointly serve all users simultaneously
without cell boundaries \cite{Interdonato2019a, CF_mimo, Nayebi2015a}.
Power control is a critical aspect of both co-located and CF Massive MIMO
systems and  has been extensively studied in many different
scenarios \cite{massivemimobook,power_control,Guo2014a, cf_maxmin,
  MMSE_LSFD, CF_MMSE} under the common assumption of an infinite
backlog, i.e., there that is an infinite amount of data waiting to be
transmitted. Since the ergodic rates of both correlated and
uncorrelated fading channels can be obtained in a tractable form in
the backlogged case, the power control can be optimized with respect
to the long-term rate performance, instead of changing with the
small-scale fading realizations as was previously common practice
\cite{Bjornson2016b}. But practical systems do not have
infinite backlogs (which would have implied infinitely long delays in
the data delivery).  Since the wireless data traffic usually arrives in
a random and bursty manner, and the packets are delivered in a few
milliseconds, the set of active users will change dynamically over
time. Considering a multi-user MIMO system with transmission queues
that contain data to be transmitted over the wireless channel, the
burstiness of data traffic becomes an important factor for optimal
resource allocation, power control and scheduling policy design
\cite{cui2010queue, dynamic_mimo, ldd_mimo}.

\subsection{Related Work}

The network throughput or spectral efficiency has always been an
important criterion to measure the capacity and efficiency of wireless
networks. Motivated by emerging delay-critical applications such as
Tactile Internet \cite{tactile_mag}, it is well understood that
delay plays an important role in the network performance
evaluation. In \cite{delay_theories}, several systematic approaches
have been listed for their ability to handle delay-aware control and
resource allocation problems. One of those approaches is Lyapunov
optimization theory, which is a powerful tool for stochastic network
optimization where dynamic control actions are made in a network to
ensure system stability with performance optimization.
	
The theory of Lyapunov drift and Lyapunov optimization is presented in
detail in \cite{neely2010stochastic}, with many examples of its
applications to communication and queueing systems. Stochastic control
for heterogeneous networks with time-varying channels has been studied
in \cite{fairness_neely}, where the optimal control strategy is
decoupled into subproblems of flow control, routing and resource
allocation. The proposed algorithms based on the drift-plus-penalty
(DPP) technique are shown to provide stability and achieve
time-average throughput arbitrarily close to the optimal fairness
operating point. Similar types of DPP-based algorithms can be found in
\cite{dynamic, utility_optimal}. A general presentation of cross-layer
control (CLC) and resource allocation strategies can be found in
\cite{georgiadis2006resource}, with special focus on flow control
algorithms that achieve optimal network fairness with stability
guarantees. Recently, Lyapunov optimization has been used to study
power control and scheduling in delay-aware device-to-device
communication \cite{d2d} and packet-based communication with deadlines
\cite{deadline}.
	
Although the use of Lyapunov optimization in communication systems is
not a new topic, very few works have considered its application to
multi-user MIMO or Massive MIMO systems. As a result of the spatial
multiplexing, the transmission queues of different users are coupled,
which adds difficulty in the maximum weighted sum rate problem that
arises when using Lyapunov optimization techniques.  MIMO downlink
scheduling with imperfect channel state information using the flow
control algorithm was studied in \cite{scheduling}, where the backlog
at the BS is assumed to be infinite. Due to the difficulty in solving
the weighted sum rate maximization problem, an on-off scheduling
policy was considered as an approximation of the optimal rate
allocation problem. The problem of ultra-reliable and low-latency
communication in millimeter wave-enabled Massive MIMO systems was
studied in \cite{urllc}, where a dynamic scheduling scheme was
developed with latency constraints, after making some simplifying
assumptions on the transmission rate expressions. Note that
\cite{power_control} proposes an effective algorithm to solve the
weighted sum rate maximization problem for single-cell co-located
Massive MIMO systems with i.i.d. Rayleigh fading channels, which
facilitates the application of Lyapunov optimization in Massive MIMO
with random data traffic.
	
\subsection{Contributions}

In this work, we develop a dynamic scheduling algorithm (DSA) that
combines cross-layer flow control with dynamic rate allocation for the
Massive MIMO uplink with randomly arriving traffic. This is among the
first ones that considers the impact of bursty traffic on the power
allocation in Massive MIMO. The algorithm decides for each time slot
on the amount of data that can be admitted to the transmission queues
and allocates appropriate transmission rates to each user.  Using
Lyapunov optimization theory, our dynamic control policy stabilizes
all transmission queues, while maximizing a concave non-decreasing
fairness function on the long-term user throughput. We show that our
DSA can greatly reduce the time-average delay of the network,
especially in cases when the optimal transmission rates derived
for saturated users cannot stabilize the queues.
	
Compared to the conference version of this paper \cite{ICC}, which
considers only co-located Massive MIMO, we have added substantial new
content about dynamic control and optimization in CF Massive MIMO,
where the maximum weighted sum rate problem is solved by using the
weighted minimum mean-squared error (MMSE) method. To make the paper
self-contained, we have also added brief explanations about the
implementation of several baseline heuristic power control algorithms.
In addition to maximum ratio combining (MRC) considered in \cite{ICC},
we also show the performance of the algorithm when using zero-forcing
(ZF), and with more fairness utility functions.

\section{Network Model}
We consider two types of uplink Massive MIMO systems, depending on
whether the antennas are co-located or distributed at different
locations. The first is a single-cell network where a base station
(BS) with $M$ co-located antennas serves $K$ single-antenna users
simultaneously. The second is a CF Massive MIMO network where $M$
single-antenna access points (APs) are scattered within a large
geographical area, and they are connected to a centralized CPU unit
for data encoding and decoding. In this work, the main difference
between these two types of networks is the achievable ergodic rates
under a given power control scheme, which will be presented in
Section~\ref{sec:rates}.

We assume that the transmission time of the physical layer (PHY) data is divided into fixed-size slots, where each slot contains the transmission time of one or multiple PHY frames.
At each time slot $t\in \{0,1,2,\ldots\}$, uplink data packets from user $k$ are generated according to a stationary and ergodic stochastic process $B_{k}(t)$ and the data generation/arrival rate of this user is $\lambda_k=\mathbb{E}[B_{k}(t)]$ bit/slot. The packet-generating processes for the $K$ users are independent of each other.
The generated data is stored in the transport layer reservoir, which is assumed to have infinite size.\footnote{This work can be easily extended to the case with a finite-size reservoir. The only difference is that the transport layer reservoir are updated as $	L_k(t+1)=\min\{\max[L_k(t)-A_k(t),0]+B_k(t), L_{\max}\}$, where $L_{\max}$ is the size of the reservoir. Upon each data arrival/generation, all data that does not fit in the reservoir will be dropped. } Similar to \cite{queue_adaptive}, we assume that each user maintains a transmission queue at the data link layer, which contains the data ready to be transmitted over the wireless channel to the BS. Denote by $L_k(t)$ and $Q_k(t)$ the amount of data (in bits) in the transport layer reservoir and in the transmission queue of user $k$ at slot $t$, respectively. To avoid congestion in the transmission queues, only a fraction of the data in the reservoir is allowed to enter the transmission queue.\footnote{The separation between transport layer reservoir and PHY transmission queues also helps us to use Lyapunov optimization framework, since it requires stable queues.} The amount of admitted data at each slot $t$ is denoted by $A_k(t)$ with $r_k=\lim\limits_{t\rightarrow\infty}\frac{1}{t}\sum\limits_{\tau=0}^{t-1}\mathbb{E}[A_k(\tau)]$ bit/slot being the time-average admitted date rate of user $k$. 
Due to the random data arrivals, the admitted data to the transmission queues at every slot $t$ must not exceed the total amount of data in the reservoir: $A_k(t)\leq L_k(t)$.

Denote by $R_k(t)$ the instantaneous PHY transmission rate of uplink user $k$ in slot $t$, measured by the number of bits that can be delivered over the wireless channel to the BS. The transmission queue $Q_k(t)$ is updated by the following equation:
\begin{equation}
Q_k(t+1)=\max[Q_k(t)-R_k(t),0]+A_k(t), \quad \forall k.
\label{eq:evolution_Q}
\end{equation}
Here, the transmission rates $R_k(t)$ are limited by the network topology, the channel statistics, and the power constraints.

\begin{figure}[ht!]
	\centering
	\includegraphics[width=1\columnwidth]{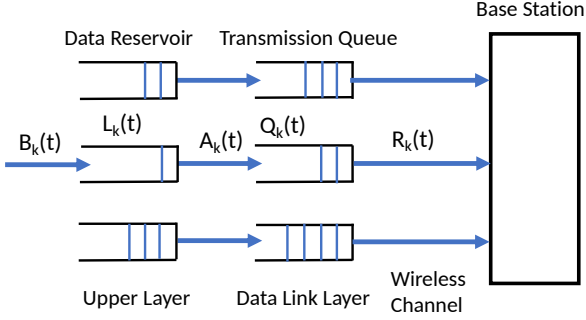}
	\caption{The structure of uplink Massive MIMO system, which consists of the data backlog reservoir and the transmission queues. }
	\label{fig:system}
\end{figure}

\begin{figure}[ht!]
	\centering 
	\includegraphics[width=1\columnwidth]{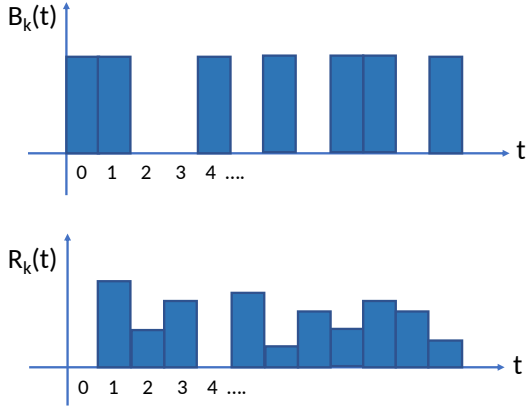} 
	\caption{An illustration example for the processes of data arrivals and PHY transmissions. At each slot, a certain amount of data is generated at each user according to some probability distribution. As a result of the dynamic power control, the PHY transmission rates might vary in different slots.}
	\label{fig:arrival_transmission}
\end{figure}

For notational convenience, we define the queue vectors $\mathbf{Q}(t)=[Q_1(t),\ldots, Q_K(t)]$, $\mathbf{R}(t)=[R_1(t),\ldots, R_K(t)]$, $\mathbf{A}(t)=[A_1(t),\ldots, A_K(t)]$ and $\mathbf{L}(t)=[L_1(t),\ldots, L_K(t)]$.
The system model is shown in Fig.~\ref{fig:system}. The evolution of the data arrival and PHY transmission processes is illustrated in Fig.~\ref{fig:arrival_transmission}.

\subsection{Problem Formulation}
Due to the random data arrivals, the $K$ uplink users in the network will not always have data to transmit.
The long-term throughput of the network, which is defined by the successful data delivery rate, will be limited by the generated data rates $\boldsymbol\lambda=[\lambda_1,\ldots,\lambda_K]$.

For arbitrary data arrival rates, our objective is to develop a dynamic control policy that:
\begin{enumerate}
	\item maintains the transmission queues stable;
	\item achieves a long-term throughput vector that maximizes some utility function $f(\cdot)$.
\end{enumerate}

In this dynamic control problem, at every slot $t$, we need to decide the amount of data to be admitted to the transmission queues, and perform power control that determines the transmission rates allocated to each user. Thus, the control decisions are $\boldsymbol{\alpha}(t)=[\mathbf{A}(t); \mathbf{p}(t)]$, where $\mathbf{p}(t)=[p_1(t), \ldots, p_K(t)]$ is the  power control vector. Let $\mathcal{A}(t)$ represent the set of all possible control decisions in slot $t$, given the random data arrivals and power constrains in that slot.

When the transmission queues are stable, the long-term throughput vector is equal to the time-average admitted data rate vector $\boldsymbol{r}=[r_1, \ldots, r_K]$. We define $\overline{X}=\lim\limits_{t\rightarrow\infty}\frac{1}{t}\sum\limits_{\tau=0}^{t-1}\mathbb{E}\{X(\tau)\}$ as the time-average of a random process $X(t)$. The utility maximization is thus defined by the following stochastic optimization problem:
\begin{subequations}
	\begin{align}
	\maximize~~&f(\boldsymbol{r}) \label{eq:optimization_utility}\\
	\textrm{subject~to}~~&0\leq r_{k}\leq \lambda_k, \quad \forall k \label{condi2}\\
	&\overline{Q}_k< \infty, \quad \forall k \label{condi3}\\
	&\boldsymbol{\alpha}(t)\in\mathcal{A}(t), \quad \forall t.
	\end{align}
	\label{eq:optimization-prob}
\end{subequations}
Here, the network utility function $f(\cdot)$ needs to be an element-wise non-decreasing concave function. It can reflect one out of the many fairness criteria that will be presented in Section \ref{sec:utility}.
The inequality in \eqref{condi2} ensures that the time-average throughput of user $k$ is not larger than the generated uplink data rate of this user. \eqref{condi3} is the strong stability condition of the transmission queues.

\subsection{Ergodic Rates in Massive MIMO}
\label{sec:rates}
\subsubsection{Co-located Massive MIMO}
We consider block fading channels with i.i.d.~Rayleigh fading between the $M$ BS antennas and the $K$ single-antenna users. This assumption allows us to derive a simple yet rigorous lower bound on the achievable ergodic rates of the users, which only depends on the large-scale fading parameters and the power control scheme. Though in reality the channels are unlikely to be i.i.d.~Rayleigh \cite{massivemimobook}, it has been shown in \cite{measured} that the achievable rates obtained by real measured channels are close to the rates obtained by assuming i.i.d.~Rayleigh channels. This is a consequence of the channel hardening and favorable propagation properties of Massive MIMO, which make the rates less dependent on the actual channel distributions, and mainly a functions of the average pathlosses.

The maximum achievable ergodic uplink rate of user $k$, measured in bit/slot, is lower-bounded by \cite{power_control}\footnote{To achieve the ergodic rate, we need to transmit codewords that span many channel realizations. In practice, this means transmitting at least 1\,kB of data \cite{Bjornson2016b}, which is easily done over a short time slot by using many sub-carriers.}
\begin{equation}
R_k=(\tau_c-\tau_p)\log_2(1+\text{SINR}_k),
\label{eq:rate_bit}
\end{equation}
where $\tau_c$ is the length of the coherence block and $\tau_p\in[K, \tau_c]$ denotes the length of the pilot signal. With MRC, we have
\begin{equation}
\text{SINR}_k=\frac{M p_{\text{d},k}\gamma_{k}}{1+\sum_{j=1}^{K}\beta_{j}p_{\text{d},j}}, 
\label{eq:mrc}
\end{equation}
where $\beta_k$ is the large-scale fading coefficient of user $k$, including the pathloss and shadowing; $\gamma_k=\frac{\tau_p p_{\text{p},k}\beta_{k}^2}{1+\tau_p p_{\text{p},k}\beta_{k}}$ is the mean square of the channel estimates; $p_{\text{p},k}$ and $p_{\text{d},k}$ denote the pilot and payload power levels, respectively.

With ZF, we have
\begin{equation}
\text{SINR}_k=\frac{(M-K) p_{\text{d},k}\gamma_{k}}{1+\sum_{j=1}^{K}p_{\text{d},j}(\beta_{j}-\gamma_{j})}.
\label{eq:zf}
\end{equation}
More details on the ergodic rates and their derivation can be found in \cite[Chapter 3]{marzetta2016fundamentals} and \cite{massivemimobook}. Since the transmission rate is not necessarily an integer, we assume that the data can be admitted and transmitted as fractional frames.

\subsubsection{CF Massive MIMO}
We consider using the large scale fading decoding (LSFD) receivers in CF Massive MIMO, where each AP computes its local estimates of the received data using a local combining vector, then passes them to the CPU. The estimate of the transmitted data is then obtained by linearly combining all received local estimates using the LSFD vectors \cite{MMSE_LSFD, CF_MMSE}.

Denote by $g_{mk}$ the channel gain between the $m$-th AP and $k$-th user. The channel is modeled by $g_{mk}\sim\mathcal{CN}(0, \beta_{mk})$, where $\beta_{mk}$ represents the large-scale fading coefficient. During the data transmission phase, the received signal at the $m$-th AP is given by
\begin{equation}
y_m=\sum\limits_{k=1}^{K}\sqrt{p_{\text{d},k}} g_{mk}s_k+n_k,
\end{equation}
where $n_k\sim \mathcal{CN}(0,1)$. The $m$-th AP obtains a local estimate $\tilde{s}_{mk}$ of the data symbol transmitted from user $k$ by using a linear decoder $v_{mk}$, i.e., $\tilde{s}_{mk}=v_{mk}y_{m}$. When using MRC, $\tilde{s}_{mk}=\hat{g}_{mk}y_{m}$, where $\hat{g}_{mk}$ is the MMSE estimate of $g_{mk}$. When $\tau_p\geq K$, we can assign pairwise orthogonal pilot sequence to each user. It has been shown that $\hat{g}_{mk}\sim\mathcal{CN}\left(0, \gamma_{mk}\right)$, with $\gamma_{mk}=\frac{\tau_p p_{\text{p},k}\beta_{mk}^2}{1+\tau_p p_{\text{p},k}\beta_{mk}}$ \cite{CF_mimo}.

After the local processing, the CPU performs a second layer decoding using the LSFD vector $\mathbf{a}_k=[a_{1k},\ldots,a_{Mk}]^{T}$. The estimate of data symbol $s_k$ is obtained by
\begin{equation}
\hat{s}_k=\mathbf{a}_k^{H}\tilde{\mathbf{s}}_{k},
\end{equation}
where $\tilde{\mathbf{s}}_{k}=[\tilde{s}_{1k},\ldots, \tilde{s}_{Mk}]^T$.

The achievable rate of user $k$ is $R_k=(\tau_c-\tau_p)\log_2(1+\text{SINR}_k)$, where $\text{SINR}_k$ is given by
\begin{align}
\text{SINR}_k&=\frac{p_{\text{d},k} \left(\sum\limits_{m=1}^{M}a_{mk}\gamma_{mk}\right)^2}{\sum\limits_{i=1}^{K}p_{d,i}\sum\limits_{m=1}^{M}a_{mk}^2 \gamma_{mk}\beta_{mi}+\sum\limits_{m=1}^{M}a_{mk}^2\gamma_{mk}}.
\label{eq:sinr_cf}
\end{align}

Note that the maximum achievable ergodic rates presented in this section are obtained when assuming that all users have saturated queues. Let $\mathbf{R}^{*}=[R_1^*, \ldots, R_K^*]$ represent the optimal transmission rates with saturated users under a certain fairness policy. 
With random data traffic, if we apply the same deterministic power control scheme, the delivered rate of user $k$ in any slot $t$ will be bounded by $R_k(t)\leq R_k^*$. In some cases, such power allocation schemes will not be able to stabilize the network even when the data arrival rates are within the network capacity region $\boldsymbol{\Lambda}$. However, using Lyapunov optimization framework, we can always achieve system stability whenever $\boldsymbol{\lambda}\in \boldsymbol{\Lambda}$, which is the main motivation of this work.

\section{Lyapunov Optimization and Dynamic Joint Scheduling and Power Control Algorithm}
For a general stochastic optimization problem, we can apply the ``min drift-plus-penalty" technique to develop a dynamic scheduling algorithm that achieves arbitrarily close performance to the optimal solution \cite{neely2010stochastic}. Recall that the original problem defined in \eqref{eq:optimization-prob} involves maximizing a concave function of a time-average quantity.  Therefore, we introduce auxiliary variables $\boldsymbol{\nu}(t)=[\nu_{1}(t),\ldots,\nu_{K}(t)]$  for each admitted data stream $A_{k}(t)$ and the corresponding virtual queues $\mathbf{Y}(t)=[Y_1(t),\ldots, Y_K(t)]$ that evolves as follows\footnote{Note that with arbitrary data arrivals, if the utility function $f(\cdot)$ is linear, maximizing $\overline{f(\boldsymbol{A})}$ is equivalent to maximizing $f(\boldsymbol{r})$. We can use a CLC1-type algorithm as proposed in \cite{fairness_neely} to achieve stability with performance optimization. For general utility functions which are not necessarily linear, maximizing functions of time-average utility requires the usage of auxiliary variables and virtual queues, as explained in \cite{neely2010stochastic}.}
\begin{equation}
Y_k(t+1)=\max[Y_k(t)-A_k(t),0]+\nu_{k}(t).
\label{eq:evolution_Y}
\end{equation}
Here, $\nu_{k}(t)$ and $Y_k(t)$ are also measured in bits.

\begin{lemma}
	The original problem in \eqref{eq:optimization-prob} which involves optimizing functions of time averages can be transformed into the following problem, which only involves time averages:
	\begin{subequations}
		\begin{align}
		\maximize~~&\overline{f(\boldsymbol{\nu})}\\
		\textnormal{subject~to}~~& \overline{\nu}_k\leq r_k, \quad \forall k \label{cons1}\\
		&\overline{Q}_k<\infty, \quad \forall k\\
		& 0\leq \nu_k(t)\leq A_{\max}, \quad \forall k, t \label{cons2}\\
		&\boldsymbol{\alpha}(t)\in\mathcal{A}(t), \quad \forall t.
		\end{align}
		\label{eq:optimization-prob2}
	\end{subequations} $\overline{f(\boldsymbol{\nu})}=\lim\limits_{t\rightarrow\infty}\frac{1}{t}\sum\limits_{\tau=0}^{t-1}\mathbb{E}[f(\boldsymbol{\nu}(\tau))]$ and $\overline{\nu}_k=\lim\limits_{t\rightarrow\infty}\frac{1}{t}\sum\limits_{\tau=0}^{t-1}\mathbb{E}[\nu_k(\tau)]$ 
	are the time-average of the utility function and the arrival rate of the virtual queue, respectively. $A_{\max}$ serves as an upper bound for the auxiliary variables and it is chosen to be suitably large such that $A_{\max}\geq r_{k}$ always holds.
\end{lemma}
\begin{IEEEproof}
	The proof to show that these two problems are equivalent can be found in \cite{neely2010stochastic} and \cite{georgiadis2006resource}.
\end{IEEEproof}

Let $\boldsymbol{\Theta}(t)=[\mathbf{Y}(t),\mathbf{Q}(t)]$ denote the combined vector of virtual queues and transmission queues. We consider the following quadratic Lyapunov function
\begin{equation}
\mathcal{L}(\boldsymbol{\Theta}(t))=\frac{1}{2}\sum\limits_{k=1}^{K}Q_k^2(t)+\frac{\eta}{2}\sum\limits_{k=1}^{K}Y_k^2(t),
\label{eq:lyapounov-function}
\end{equation}
where $0<\eta\leq 1$ is a bias factor that determines the relative weight on the virtual queues. The Lyapunov function is a scalar measure of the congestion level in the system.
The one-step conditional Lyapunov drift is defined by\footnote{The Lyapunov drift measures the difference between the network congestion level in two consecutive slots. Intuitively, if the drift is minimized at every slot, meaning that the congestion will be reduced progressively, the queues will eventually be stabilized.}
\begin{equation}
\Delta\big(\boldsymbol{\Theta}(t)\big)=\mathbb{E}[\mathcal{L}(\boldsymbol{\Theta}(t+1))-\mathcal{L}(\boldsymbol{\Theta}(t))|\boldsymbol{\Theta}(t)],
\label{eq:one-step-drift}
\end{equation}  
where the expectation is with respect to the random data arrivals and possible control actions.
In order to stabilize the system while optimizing the network utility, at every time slot, we minimize a bound on the following drift-plus-penalty metric\footnote{The basic philosophy of the min-drift-plus-penalty technique is to make control decisions that balances the network congestion and the optimal network utility.}
\begin{equation}
\Delta(\boldsymbol{\Theta}(t))-V \mathbb{E}[f(\boldsymbol{\nu}(t))|\boldsymbol{\Theta}(t)],
\end{equation}
where $V$ is a control parameter that leverages between improved network utility and increased congestion in the queues.

\begin{lemma}
	\label{lemma2}
	The drift-plus-penalty is upper bounded as 
	\begin{equation} 
	\begin{split}
	&\Delta(\boldsymbol{\Theta}(t))-V \mathbb{E}[f(\boldsymbol{\nu}(t))|\boldsymbol{\Theta}(t)]\\
	\leq  
	&~C-\mathbb{E}\left[\sum_{k=1}^{K}A_k(t)\big(\eta Y_k(t)-Q_k(t)\big)|\boldsymbol{\Theta}(t)\right]  \\
	&-\mathbb{E}\left[Vf(\boldsymbol{\nu}(t))\!-\!\eta \sum\limits_{k=1}^{K}Y_k(t)\nu_k(t)|\boldsymbol{\Theta}(t)\right]\\
	&-\mathbb{E}\left[\sum_{k=1}^{K}Q_k(t) R_k(t)|\boldsymbol{\Theta}(t)\right],
	\label{eq:drift-penalty}
	\end{split}
	\end{equation}
	\vspace{-0.1cm}
	where $C=\frac{1}{2}\sum\limits_{k=1}^{K}R_{k,\max}^2+\frac{2\eta+1}{2}K A_{\max}^2$ and $R_{k,\max}$ is the maximum achievable rate of user $k$ when only user $k$ is scheduled to transmit.
\end{lemma}
\begin{IEEEproof}
	See Appendix \ref{appen1}.
\end{IEEEproof}

Using the notion of opportunistically minimizing a conditional expectation, minimizing the drift-plus-penalty bound in \eqref{eq:drift-penalty} at every slot $t$ leads to the need to solve three subproblems, which are solved in the following subsections.

\subsection{First Subproblem: Data Admission Control}
To minimize the first non-constant term at the right-hand side of \eqref{eq:drift-penalty}, we need to choose $A_k(t)$ that maximizes $\sum_{k=1}^{K}A_k(t)\big(\eta Y_k(t)-Q_k(t)\big)$ under the backlog constraint $A_k(t)\leq L_k(t)$ and the admission burst limit $0\leq A_k(t)\leq A_{\max}$. The solution to the first subproblem is
\begin{equation}
A_k(t)= 
\left\lbrace 
\begin{array}{ccc}
\!\!\min\{L_k(t), A_{\max}\}
& \text{if}~Q_k(t)\leq\eta Y_k(t), \\
0
& \text{otherwise.}
\end{array} \right.
\end{equation}

\subsection{Second Subproblem: Auxiliary Variables}	
\label{sec:utility}
To minimize the second non-constant term of the right-hand side of \eqref{eq:drift-penalty}, we need to choose the auxiliary variables $0\leq \nu_k(t)\leq A_{\max}$ that solve
\begin{equation}
\maximize\limits_{\substack{0\leq\nu_k(t)\leq A_{\max}\\\boldsymbol{\nu}(t)=[\nu_1(t),\ldots,\nu_{K}(t)]}}~Vf(\boldsymbol{\nu}(t))-\eta \sum\limits_{k=1}^{K}Y_k(t)\nu_k(t). 
\label{eq:utility-maximization}
\end{equation}
The solution to this problem depends on the specific utility function. Three examples are given below. 

\subsubsection{Max-Min Fairness (MMF)}
In this case, every user should achieve the same performance, thus we have the utility function 
\begin{equation}
f(\boldsymbol{\nu}(t))=\min\{\nu_1(t), \nu_2(t),\ldots, \nu_K(t)\}.
\end{equation}
The solution to \eqref{eq:utility-maximization} is the case when all $\nu_k(t)$ are the same and when $\nu_k(t)\left[V-\eta \sum_{j=1}^{K}Y_j(t)\right]$ is maximized. Combined with $0\leq \nu_k(t)\leq A_{\max}$, we have the solution to \eqref{eq:utility-maximization} as
\begin{align}
\nu_k(t)= \left\{
\begin{array}{rcl}
&A_{\max} & \text{if}~~V>\eta\sum_{j=1}^{K}Y_j(t),\\
&0 & \text{otherwise.}\\
\end{array} \right.
\end{align}

\subsubsection{Proportional Fairness (PF)}
In a multi-user system, a scheduler is said to achieve PF if the long-term throughput vector $[r_1,\ldots,r_K]$ maximizes $\sum_{k=1}^{K}\log r_k$. It is equivalent to maximizing the geometric mean of the rates. 
We have the utility function as
\begin{equation}
f(\boldsymbol{\nu}(t))=\sum\limits_{k=1}^{K}\log (\nu_k(t)).
\end{equation}
Plugging it into \eqref{eq:utility-maximization}, we search for $\nu_{k}(t)$ that maximizes $g(\boldsymbol{\nu}(t))=V\sum\limits_{k=1}^{K}\log (\nu_k(t))-\eta \sum\limits_{k=1}^{K}Y_k(t)\nu_k(t)$. Taking the first order derivative of $g(\boldsymbol{\nu}(t))$over $\nu_k(t)$ yields
\begin{equation} 
\frac{\partial g(\boldsymbol{\nu}(t))}{\partial \nu_k(t)}=\frac{V}{\nu_k(t)}-\eta Y_k(t).
\end{equation}
Since the second-order derivative is strictly negative, $g(\boldsymbol{\nu}(t))$ is a concave function of $\nu_{k}(t)$. The maximum point is achieved when $\frac{\partial g(\boldsymbol{\nu}(t))}{\partial \nu_k(t)}=0$, which yields $\nu_{k}(t)=\frac{V}{\eta Y_k(t)}$. Combined with the constraint $0\leq\nu_k(t)\leq A_{\max}$, the optimal solution is given by
\begin{equation}
\nu_k(t)=\min\left\{\frac{V}{\eta Y_k(t)}, A_{\max}\right\}.
\end{equation}

\subsubsection{Maximum Sum Rate (MSR)}
In this case, the system should maximize the data throughput without taking fairness between users into consideration. We then have the utility function
\begin{equation}
f(\boldsymbol{\nu}(t))=\sum_{k=1}^{K} \nu_k(t).
\end{equation}
The solution to \eqref{eq:utility-maximization} is
\begin{align}
\nu_k(t)= \left\{
\begin{array}{rcl}
&A_{\max} & \text{if}~~V>\eta Y_k(t),\\
&0 & \text{otherwise.}\\
\end{array} \right.
\end{align}

In the aforementioned three cases, we are able to derive the closed-form expressions of the auxiliary variables $\nu_k(t)$ that solve \eqref{eq:utility-maximization}. For other concave non-decreasing utility functions, the closed-form solutions might not be available, standard convex solvers can be used to find the optimal solutions. Note that we can introduce some constant weights into the fairness criteria above and some problems might still be solved in closed-form.

\subsection{Third Subproblem: PHY Rate Allocation} 
To minimize the third non-constant term at the right-hand side of \eqref{eq:drift-penalty}, we choose $R_k(t)$ that maximizes $\sum_{k=1}^{K}Q_k(t) R_k(t)$, which we identify as a weighted sum rate problem.
This is often the most challenging subproblem since the maximization of the weighted sum rate is an NP hard problem in many cases \cite{Luo2008a}.

\subsubsection{Co-located Massive MIMO}
Recently, \cite{power_control} developed an efficient algorithm that exploits the special structure of the rates in Massive MIMO to solve this weighted sum rate optimization problem. The algorithm is briefly explained as follows. 

Assuming full power for pilot transmission, we have $\gamma_{k}=\frac{P_{\max}\tau_p \beta_{k}^2}{1+P_{\max}\tau_p \beta_{k}}$. For the case with MRC, the optimization problem is given by
\begin{subequations}
	\begin{align}
	\maximize\limits_{\{p_k\}}~~&\sum_k Q_k\log\left(1+\frac{M p_{k}\gamma_{k}}{1+\sum_{j=1}^{K}\beta_{j}p_{j}}\right)\\
	\textnormal{subject~to}~~& 0\leq p_k\leq P_{\max}, \quad \forall k.
	\end{align}
\end{subequations}
Noticing that the all the users have the same denominator in the SINR expression, the original problem can be transformed into the following problem:
\begin{subequations}
	\label{eq:opt_weighted_sum}
	\begin{align}
	\maximize\limits_{s,\{x_k\}}~~&\sum_k Q_k\log\left(1+a_k x_k\right) \\
	\textnormal{subject~to}~~& 0\leq x_k\leq \beta_k P_{\max} s, \quad \forall k\\
	& \sum_{j=1}^{K}x_j=1-s,
	\end{align}
\end{subequations}
where $a_k=\frac{M \gamma_k}{\beta_k}$ with MRC and $a_k=\frac{(M-K) \gamma_k}{\beta_k-\gamma_{k}}$ with ZF. 
Since the new problem is convex, standard convex solvers can be used to find the optimal solution.

\subsubsection{CF Massive MIMO}
Let $p_{\text{d},k}=P_{\max} \eta_k^2$, where $P_{\max}$ is the maximum transmission power and $\eta_k^2$ is the power control coefficient of user $k$.\footnote{Here, we use $\eta_k^2$ instead of $\eta_k$, in order to simply the expressions when using the weighted MMSE method.} 
To get a local optimum point for the maximum weighted sum rate problem $\sum\limits_{k=1}^{K}Q_k R_k$, we use the weighted MMSE method as in \cite{LSFD}, which gives us the following iterative algorithm.

\begin{theorem}
	Let $u_k^{(n-1)}$, $e_k^{(n-1)}$, $\mathbf{a}_k^{(n-1)}$, $\eta_k^{(n-1)}$ denote the parameter values in the iteration $n-1$. At current iteration $n$, these optimization parameters are updates by the following equations:
	\begin{enumerate}
		\item Update $u_k^{(n)}$ by 
		\begin{equation}
		u_k^{(n)}=\frac{\sqrt{P_{\max}}\eta_k^{(n-1)}\sum_{m=1}^{M}a_{mk}^{(n-1)}\gamma_{mk} }{d_k^{(n)}},
		\end{equation}
		where $d_k^{(n)}$ is defined in \eqref{eq:d}, shown at the top of the next page.
		\item Update $e_k^{(n)}$ by 
		\begin{equation}
		e_k^{(n)}=\left(u_k^{(n)}\right)^2d_k^{(n)}-2\sqrt{P_{\max}}\eta_k^{(n-1)} u_k^{(n)}\sum_{m=1}^{M}a_{mk}^{(n-1)}\gamma_{mk}+1.
		\end{equation}
		Obtain $w_k^{(n)}=Q_k/e_k^{(n)}$.
		\item Update $\mathbf{a}_k^{(n)}$ by 
		\begin{equation}
		\mathbf{a}_k^{(n)}=\frac{\sqrt{P_{\max}}\eta_k^{(n-1)}}{u_k^{(n)}}C_k^{-1} \boldsymbol{\mu}_k,
		\end{equation}
		where $\boldsymbol{\mu}_k=[\gamma_{1k},\ldots,\gamma_{Mk} ]^{T}$, $C_k$ is defined in \eqref{eq:Ck}.
		\item  Update $\eta_k^{(n)}$ by $\eta_k^{(n)}=\min\{\tilde{\eta}_k^{(n)},1\}$, where $\tilde{\eta}_k^{(n)}$ is given in \eqref{eq:eta}.
	\end{enumerate}
\end{theorem}
\begin{figure*}[ht!]
	\begin{equation}
	d_k^{(n)}=P_{\max}\left(\eta_k^{(n-1)}\right)^2\left(\sum_{m=1}^{M}a_{mk}^{(n-1)}\gamma_{mk}\right)^2+\sum_{i=1}^{K}P_{\max}\left(\eta_i^{(n-1)}\right)^2\sum_{m=1}^{M}\left(a_{mk}^{(n-1)}\right)^2 \gamma_{mk}\beta_{mi}+\sum_{m=1}^{M}\left(a_{mk}^{(n-1)}\right)^2\gamma_{mk}.
	\label{eq:d}
	\end{equation}
	\begin{equation}
	C_k=P_{\max}\left(\eta_k^{(n-1)}\right)^2 \boldsymbol{\mu}_k \boldsymbol{\mu}_k^{H}+\textnormal{diag} \left(\sum\limits_{i=1}^{K}P_{\max}\left(\eta_i^{(n-1)}\right)^2 \gamma_{mk}\beta_{mi}+\gamma_{mk}\right)_{1\leq m\leq M}.
	\label{eq:Ck}
	\end{equation}
	\begin{equation}
	\tilde{\eta}_k^{(n)}=\frac{w_k^{(n)} u_k^{(n)}\sum\limits_{m=1}^{M}a_{mk}^{(n)}\gamma_{mk}}{\sqrt{P_{\max}}\left[\left(u_k^{(n)}\right)^2  w_k^{(n)}\left(\sum\limits_{m=1}^{M}a_{mk}^{(n)}\gamma_{mk}\right)^2\!\!+\!\sum\limits_{i=1}^{K} \sum\limits_{m=1}^{M}w_i^{(n)} \left(u_i^{(n)}\right)^2\left(a_{mi}^{(n)}\right)^2\gamma_{mi}\beta_{mk}\right]}.
	\label{eq:eta}
	\end{equation}
\end{figure*}
It was proved in \cite{LSFD} that this iterative algorithm converges to a sub-optimal stationary solution to the maximum weighted sum rate problem.

In summary, we have developed the dynamic scheduling algorithm (DSA) for Massive MIMO uplink that is given in Algorithm~\ref{algorithm:DSA}, where one iteration is taken per time slot.

\begin{algorithm}
	\caption{Dynamic Scheduling Algorithm (DSA)} \label{algorithm:DSA}
	\begin{enumerate}
		\item Initialization: $L_k(0)=0$, $Q_k(0)=0$ and $Y_k(0)=0$ for all $k=1,\ldots,K$. Set $t=1$.
		\item At current slot $t$, each user $k$ observes the virtual queue $Y_k(t)$ and calculates the input $\nu_k(t)$ at the virtual queues by solving 
		\begin{equation}
		\max\limits_{\substack{0\leq\nu_k(t)\leq A_{\max}\\\boldsymbol{\nu}(t)=[\nu_1(t),\ldots,\nu_{K}(t)]}}\left[V\cdot f\big(\boldsymbol{\nu}(t)\big)-\eta\sum\limits_{k=1}^{K}Y_k(t)\nu_{k}(t)\right].
		\label{eq:solution} \nonumber
		\end{equation} 
		Here, $A_{\max}$ and $V$ are suitably large constant parameters. 
		\item Each user $k$ observes $Y_k(t)$ and $Q_k(t)$, and computes the admitted data $A_k(t)$ by 
		\begin{equation}
		A_k(t)= 
		\left\lbrace 
		\begin{array}{ccc}
		\!\!\min\{L_k(t), A_{\max}\}
		& \text{if}~Q_k(t)\leq\eta Y_k(t), \\
		0
		& \text{otherwise}.
		\end{array} \right.\nonumber
		\end{equation}
		\item The network center observes the transmission queue size vector $\mathbf{Q}(t)$, and determines the power control vector $\mathbf{p}(t)$ by solving the weighted sum rate maximization problem:
		\begin{equation}
		\max\limits_{\substack{0\leq p_k(t)\leq P_{\max}\\\mathbf{p}(t)=[p_1(t), \ldots, p_K(t)]}}~\sum_{k=1}^{K} Q_k(t)\cdot R_k(t). \label{eq:scheduler} \nonumber
		\end{equation}
		\item Update the virtual queues, transmission queues and the transport layer reservoirs at each user as follows:
		\begin{align} \notag
		Y_k(t+1)&=\max[Y_k(t)-A_k(t),0]+\nu_{k}(t),\\ \notag
		Q_k(t+1)&=\max[Q_k(t)-R_k(t),0]+A_k(t),\\ \notag
		L_k(t+1)&=\max[L_k(t)-A_k(t),0]+B_k(t).
		\end{align}
		\item Continue steps 2--5 for the next slot $t+1$.
	\end{enumerate}
\end{algorithm}

\subsection{Baseline Heuristic Schemes}
Here, we present some baseline power control schemes in the literature that can be used for comparison with our DSA. 
\subsubsection{Co-located Massive MIMO}
For performance comparison, we consider the MMF and MSR optimization algorithms proposed in \cite{power_control} and the PF optimization algorithm proposed in \cite{prop_fair_multi}. The three algorithms are briefly described as follows.
\begin{itemize}
	\item For MMF, consider full power for pilot transmission, i.e., $p_{\text{p},k}=P_{\max}$, the optimal payload data transmission power of user $k$ is $p_{d,k}=\frac{\min\{\gamma_{1},\ldots, \gamma_{K}\}}{\gamma_{k}}P_{\max}$, where $\gamma_{k}=\frac{P_{\max}\tau_p \beta_{k}^2}{1+P_{\max}\tau_p \beta_{k}}$.
	\item For PF, the optimization problem is defined by
	\begin{subequations}
		\begin{align}
		\maximize\limits_{\{\eta_k\}}~~&\sum_k\log\left(\log(1+\text{SINR}_k)\right)\\
		\textnormal{subject~to}~~& 0\leq \eta_k\leq 1, \quad \forall k. 
		\end{align}
	\end{subequations}
	Using the techniques in \cite{prop_fair_multi}, the original problem can be transformed into the following equivalent problem, for the case with MRC:
	\begin{subequations}
		\begin{align}
		\maximize\limits_{\{\eta_k\},\{t_k\}}~~&\sum_{k=1}^{K}\log\left(\log(1_{\epsilon}+e^{t_k})\right)\\
		\textnormal{subject~to}~~& e^{\eta_k}\leq 1, \quad \forall k \\
		& \frac{M\gamma_{k}P_{\max}e^{\eta_k}}{1+\sum_{j}\beta_{j}P_{\max}e^{\eta_j}}\geq e^{t_k}, \quad \forall k. \label{constraint2}
		\end{align}
	\end{subequations}
	When ZF is used, the constraint in \eqref{constraint2} becomes  $\frac{(M-K)P_{\max}e^{\eta_k}\gamma_{k}}{1+\sum_{j}(\beta_{j}-\gamma_{j})P_{\max}e^{\eta_j}}\geq e^{t_k}$. The transformed problem can be easily solved by convex solvers.
	\item For MSR, we solve the optimization problem defined in \eqref{eq:opt_weighted_sum} with $Q_k=1, \forall k$.
\end{itemize}

Since these algorithm are clearly suboptimal whenever a user has an empty queue, we consider three heuristic benchmark schemes to optimize different fairness objectives..
The first is referred to as ``modified MMF" scheme, described as follows.
\begin{enumerate}
	\item At time slot $t$, if $L_k(t)=0$, then user $k$ is removed from the list of users waiting to be served. Let $K'$ denote the number of users with non-empty queues.
	\item Apply the MMF power control algorithm from \cite{power_control} on the $K'$ users such that max-min fairness is achieved among them. In this case, all active users have the same transmission rate and some might be over-provisioned.
\end{enumerate}
Similarly, we can have ``modified MSR'' and ``modified PF'' algorithms by solving the MSR and PF optimization problems after removing empty queues at the start of each time slot.

\subsubsection{CF Massive MIMO}
A baseline power control scheme to achieve MMF  in CF Massive MIMO can be used to compare its performance with the DSA. 
The SINR expression in \eqref{eq:sinr_cf} can be rewritten as
\begin{equation}
\text{SINR}_k=\frac{P_{\max}\eta_k \mathbf{a}_k^{H}\boldsymbol{\mu}_k\boldsymbol{\mu}_k^{H}\mathbf{a}_k}{\mathbf{a}_k^{H}\boldsymbol{\Lambda}_k\mathbf{a}_k},
\end{equation}
where $\boldsymbol{\Lambda}_k=\textnormal{diag}\left(\sum\limits_{i=1}^{K}P_{\max}\eta_i \gamma_{mk}\beta_{mi}+\gamma_{mk}\right)_{1\leq m\leq M}$, and $\boldsymbol{\mu}_k=[\gamma_{1k},\ldots,\gamma_{Mk} ]^{T}$. The MMF problem is given as follows.
\begin{subequations}
	\begin{align}
	\max_{\{\mathbf{a}_k\},\{\eta_k\}} \min_{k}~~&\log(1+\text{SINR}_k)\\
	\textnormal{subject~to}~~& 0\leq \eta_k\leq 1, \quad \forall k. 
	\end{align}
\end{subequations}
Since this problem is non-convex, we use alternating optimization to develop an algorithm that solves two sub-problems and update the variables iteratively \cite{cf_maxmin}. Let $\mathbf{a}_k^{(n-1)}$, $\eta_k^{(n-1)}$ denote the values of the optimization variables in the iteration $n-1$. At current iteration $n$, these optimization parameters are updates by the following equations:
\begin{enumerate}
	\item Update $\mathbf{a}_k^{(n)}=\left(\boldsymbol{\Lambda}_k^{(n-1)}\right)^{-1}\boldsymbol{\mu}_k$, where $\boldsymbol{\Lambda}_k^{(n-1)}=\textnormal{diag}\left(\sum\limits_{i=1}^{K}P_{\max}\eta_i^{(n-1)} \gamma_{mk}\beta_{mi}+\gamma_{mk}\right)_{1\leq m\leq M}$. This is obtained by maximizing the SINR of each user.
	\item Update $\boldsymbol{\eta}^{(n)}=[\eta_1^{(n)},\ldots, \eta_K^{(n)}]$ by solving the following geometric program (GP) problem:
	\begin{subequations}
		\begin{align}
		\max_{t, \{\eta_k\}} ~~&t\\
		\textnormal{subject~to}~~& \frac{\left(\mathbf{a}_k^{(n)}\right)^{H}\boldsymbol{\Lambda}_k^{(n)}\mathbf{a}_k^{(n)}}{P_{\max}\eta_k^{(n)} \left(\mathbf{a}_k^{(n)}\right)^{H}\boldsymbol{\mu}_k \boldsymbol{\mu}_k^H\mathbf{a}_k^{(n)}}\leq \frac{1}{t},\\
		&0\leq \eta_k\leq 1, \quad \forall k 
		\end{align}
	\end{subequations}
\end{enumerate}
It was shown in \cite{cf_maxmin} that this algorithm converges to the optimal solution.

Similar to the co-located Massive MIMO case, a modified MMF algorithm can be obtained by removing users with empty queues at each slot.

\section{Performance Evaluation}
\label{sec:simulation}
In this section, we evaluate the performance of the DSA in both co-located and CF Massive MIMO networks with different utility functions, including max-min fairness (MMF), maximum sum rate (MSR) and proportional fairness (PF). 

We consider a network area of size $1\,$km$\times$$1\,$km. The number of antennas is $M=100$, and the number of users is $K=10$. For co-located Massive MIMO, the antennas are co-located at the cell center, while for the CF case, the $100$ antennas are randomly distributed in the network area. The user locations are also randomly dropped in the network area. The number of symbols per coherence interval is $\tau_c=100$, which could be achieved by having a coherence bandwidth of $B=100\,$kHz with the coherence time $T_c=1\,$ms. The pilot signal length $\tau_p$ is equal to the number of effectively scheduled users in a time slot. The payload power budget of user $k$ is $P_{\max}=10^{0.5}\times 500^{3.76}$, which gives an average SNR of $5$dB at the cell edge ($500\,$m) with normalized noise variance. The pilot power of user $k$ is set as $P_{\text{d},k}=P_{\max}$. 

For co-located Massive MIMO, the large scale fading coefficients are $\beta_{k}=z_k/r_{k}^{\alpha}$ for $k=1,\ldots,K$, where $z_k$ is the log-normal shadowing with $8$dB standard deviation and $r_{k}^{-\alpha}$ is the path loss with $\alpha=3.76$.  In the simulations, we have sorted $\{\beta_{k}\}$ such that user $10$ has the best channel condition and user $1$ has the worst. Table \ref{tab:beta} gives an example of the SNR (dB) values (in ascending order) of the $K=10$ users that we use in the simulations. For CF Massive MIMO, the large scale fading coefficients are $\beta_{mk}=\text{PL}_{mk} \cdot z_{mk}$ where $\text{PL}_{mk}$ represents the path loss and $z_{mk}$ represents the shadowing coefficient. Here, we use the same three-slope path loss model as in \cite{cf_stogeo}.

The data-generating process of user $k$ follows a memoryless Bernoulli process with packet arriving probability $p_k$ per slot, i.e., $B_k(t)\sim \text{Bernoulli}(p_k)\times B_{\max}$, with $B_{\max}$ being the packet size. The duration of one slot is the same as the length of a coherence interval.\footnote{For simplicity, we assumed that there is only one coherence interval per slot, but all the results can be readily applied to a system where one time slot contains multiple coherence intervals that are distributed over the frequency domain. The only change that is needed is to multiply $A_{\max}$, $B_{\max},$ and $V$ with the number of coherence intervals per slot.} We choose $A_{\max}=50\times \tau_c$, $V=500\times \tau_c$, $\eta=1$. The results are obtained after at least $10000$ slots.

\begin{table}[t]
	\centering
	\caption{SNR $(\textnormal{dB})$ of $K=10$ users}
	\renewcommand{\arraystretch}{1.2}
	\begin{tabular}{|c|c|c|c|c|}
		\hline   
		$\text{SNR}_1$&    $\text{SNR}_2$&  $\text{SNR}_3$&  $\text{SNR}_4$& $\text{SNR}_5$ \\
		\hline
		$ -0.62$ & $3.27$  &$ 5.4$  &$6.5$  &$ 9.5$   \\
		\hline
		$\text{SNR}_6$&    $\text{SNR}_7$&  $\text{SNR}_8$&  $\text{SNR}_9$& $\text{SNR}_{10}$ \\
		\hline
		$10$    &$12.8$  &$15.7$ &$17.56$   & $22.36$ \\
		\lasthline
	\end{tabular}
	\label{tab:beta}
\end{table}

\subsection{Throughput Comparison}
\begin{figure}[t!]
	\centering
	\includegraphics[width=0.99\columnwidth]{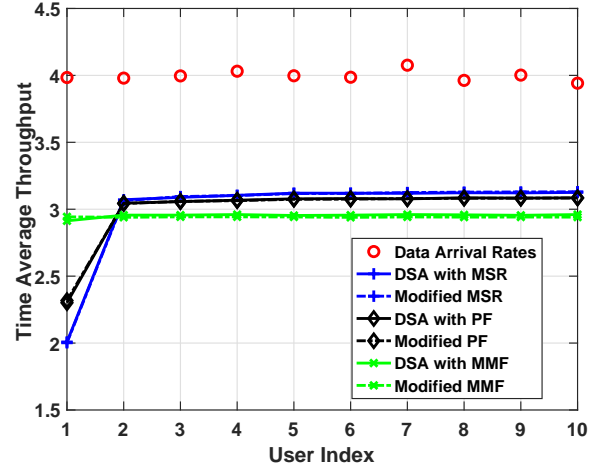}
	\caption{Time-average throughput of $K=10$ users with MRC processing in co-located Massive MIMO. $B_{\max}=4\times\tau_c$, $A_{\max}=20\times \tau_c$, $V=2000\times\tau_c$. $p_k=1, \forall k=1,\ldots,K$. }
	\label{fig:rate-one-location-mrc}
\end{figure}

\begin{figure}[t!]
	\centering
	\includegraphics[width=0.99\columnwidth]{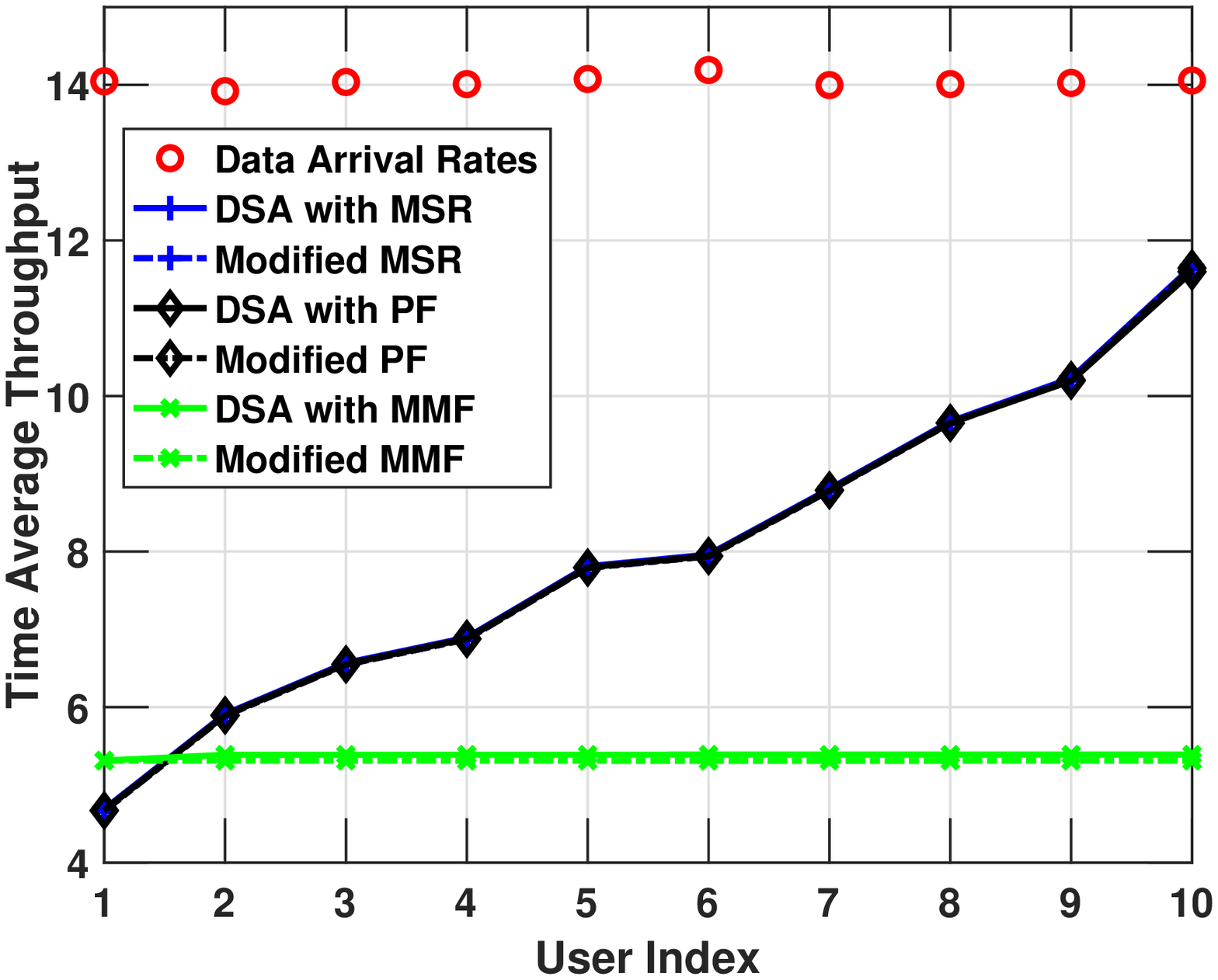}
	\caption{Time-average throughput of $K=10$ users with ZF processing in co-located Massive MIMO. $B_{\max}=14\times\tau_c$, $A_{\max}=20\times \tau_c$, $V=4000\times\tau_c$. $p_k=1, \forall k=1,\ldots,K$.}
	\label{fig:rate-one-location-zf}
\end{figure}

In Fig.~\ref{fig:rate-one-location-mrc} and Fig.~\ref{fig:rate-one-location-zf}, we compare the time-average throughput (bit/channel use) obtained by the DSA with the heuristic algorithms, when all users have high data arrival rates. The achievable ergodic rates in Step 4 of the DSA are obtained with MRC in Fig.~\ref{fig:rate-one-location-mrc} and with ZF in Fig.~\ref{fig:rate-one-location-zf}. 
In both case, the data arrival rates are outside the network capacity region, which means that the backlog goes towards infinity and the users will (almost) always have data to transmit. Hence, the time-average throughput is limited by the achievable ergodic rates of the users. 
From both figures, we see that the three heuristic algorithms provide almost the same throughput as the proposed DSA. This shows that in terms of throughput optimization, the dynamic resource allocation using Lyapunov optimization is not needed when all the users have saturated data traffic.
Note that in Fig.~\ref{fig:rate-one-location-zf}, the results obtained by ``DSA with MSR", ``Modified MSR", ``DSA with PF" and ``Modified PF" overlap with each other. More specifically, the optimal solutions for both MSR and PF correspond to the case when all users transmit with maximum power. The reason  might be that the interference is very small when ZF processing is used. Thus, both $\sum_{k=1}^{K}R_k$ and $\sum_{k=1}^{K}\log (R_k)$ are maximized when using the maximum transmit power for all users. Since the MSR and PF give very similar performance, in the remainder of this section, we only present simulation results in the cases with MMF and MSR.

\begin{figure}[t!]
	\centering
	\includegraphics[width=0.99\columnwidth]{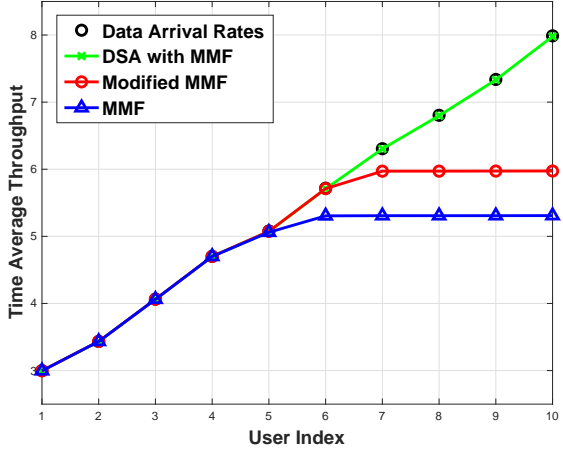}
	\caption{Time-average throughput  of $K=10$ users with ZF processing in co-located Massive MIMO. $B_{\max}=10\times\tau_c$, $p_1=0.3$, $p_{10}=0.8$, uniform spacing between $p_k$ and $p_{k+1}$ for all $k=1,\ldots, 9$.}
	\label{fig:thpt_mmf_zf_stable}
\end{figure}

Note that when the data arrival rates are outside the network capacity region, the system is not practically useful since the delays go to infinity because of unstable queues.  We therefore consider the practical case of having data arrival rates that are inside the network capacity region. In Fig.~\ref{fig:thpt_mmf_zf_stable}, we show the throughput comparison when the worst-channel user has the lowest data arrival rate. With the DSA, the throughput of all users are equal to their data arrival rates. This is because when a system is stable, the user throughput is limited by the data arrival rate. For the two baseline schemes, the throughput of some users are lower than in the DSA case, because the achievable ergodic rates of these users are smaller than their data arrival rates. This implies that the network is unstable when we use the MMF and the modified MMF schemes, even though it can potentially be stabilized.

\subsection{Delay Comparison}
\begin{figure}[t!]
	\centering
	\includegraphics[width=0.99\columnwidth]{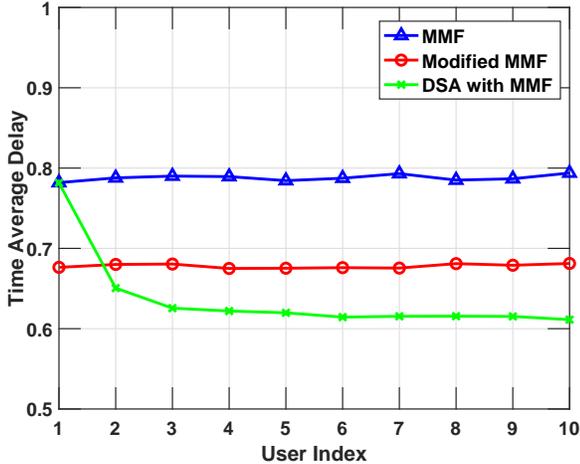}
	\caption{Time-average delay ($\times 10^{-5}\,$s/bit) of $K=10$ users with MRC in co-located Massive MIMO. All the users have the same data arrival rates, i.e., $B_{\max}=5\times\tau_c$, $p_k=0.4$ for all $k=1,\ldots, 10$.}
	\label{fig:delay_comp_mmf}
\end{figure}

In Fig.~\ref{fig:delay_comp_mmf}, we consider the time-average delay obtained with the DSA, the original MMF algorithm in \cite{power_control} and the modified MMF.  The data arrival probabilities are the same for all users.  We see that the average delays of the $10$ users are very similar when using the MMF and the modified MMF algorithms, because the allocated transmission rates are almost the same. The DSA has a clear advantage in reducing the time-average delay for most users compared to the two alternative algorithms, at the price of giving slightly larger delay for the worst-channel user.

\begin{figure}[t!]
	\centering
	\includegraphics[width=0.99\columnwidth]{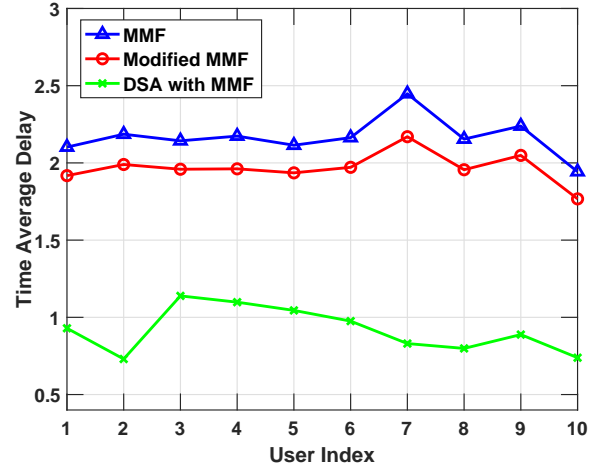}
	\caption{Time-average delay ($\times 10^{-5}\,$s/bit) of $K=10$ users with MRC  in co-located Massive MIMO. User locations change over time based on random walk model with maximum step $5\,$m. The results are obtained after $t=5\times 10^4$ slots, while user locations change every $100$ slots. Other parameters are the same as in Fig.~\ref{fig:delay_comp_mmf}. }
	\label{fig:delay_comp_mmf_randomwalk}
\end{figure}

In additional to the case with users at fixed location, we present in Fig.~\ref{fig:delay_comp_mmf_randomwalk} the time-average delay when the user locations change over time. We consider a random-walk-based process to model the mobility of users where each step is taken with random distance between $[0, 5]\,$m in a arbitrary directions.\footnote{If the new location falls outside the cell range, another random step is generated until the new location falls inside the cell.} Compared to Fig.~\ref{fig:delay_comp_mmf} obtained with fixed user locations, the delay improvement of the DSA is more profound when users are moving around. This is because the two baseline power allocation schemes only depend on the channel coefficients  at current slot, while the DSA operates in a dynamic way that could take into account the variation of the channel coefficients in future slots.

\begin{figure}[t!]
	\centering
	\includegraphics[width=0.99\columnwidth]{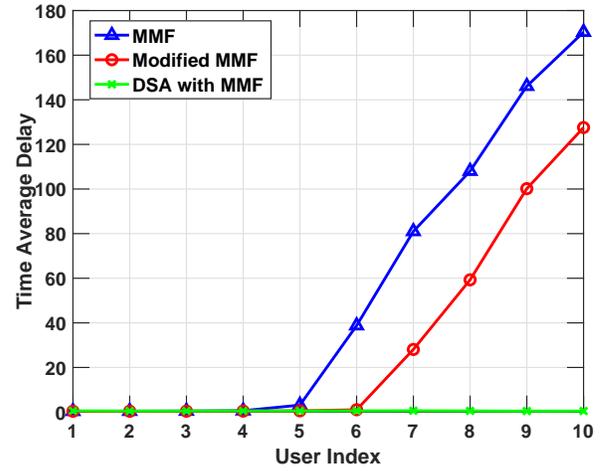}
	\caption{Time-average delay ($\times 10^{-5}\,$s/bit) of $K=10$ users with ZF  in co-located Massive MIMO. The results are obtained after $t=10^{4}$ time slots.  $B_{\max}=10\times\tau_c$, $p_1=0.3$, $p_{10}=0.8$, uniform spacing between $p_k$ and $p_{k+1}$ for all $k=1,\ldots, 9$.}
	\label{fig:delay_comp_mmf_zf}
\end{figure}
\begin{figure}[t!]
	\centering
	\includegraphics[width=0.99\columnwidth]{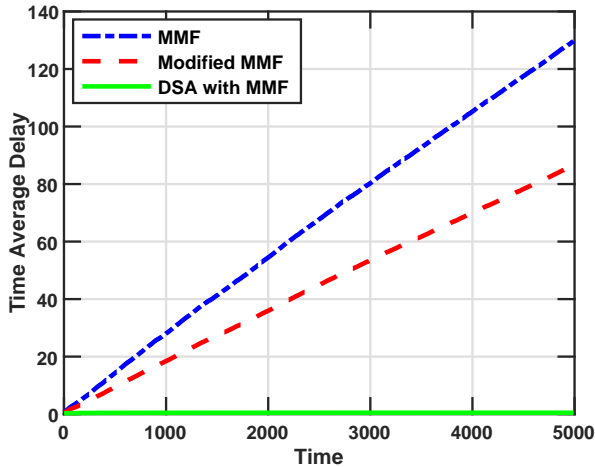}
	\caption{Time-average delay ($\times 10^{-5}\,$s/bit) of user $10$ vs. time. Same parameters as in Fig.~\ref{fig:delay_comp_mmf_zf}.}
	\label{fig:delay_increase}
\end{figure}

In Fig.~\ref{fig:delay_comp_mmf_zf}, we present the delay comparison of the MMF schemes with ZF, when users have different data arrival rates. From Fig. \ref{fig:rate-one-location-zf} we can see that the optimal MMF rate with ZF in the infinite backlog case is $R_{k}^{*}= 5.3$ bit/channel use for all $k$. The arrival rate vector we choose for Fig.~\ref{fig:delay_comp_mmf_zf} is  within the network capacity region, but some users (e.g., users 6-10)  have higher data arrival rates than $R_{k}^{*}$. It is expected that the MMF scheme will not be able to stabilize the network. As we can see from the figure, the DSA gives very small and balanced delay for all users, while for the two baseline schemes, some users will have infinite delay.\footnote{Note that when the queue is unstable, the time-average delay is infinite. The delay values presented in Fig.~\ref{fig:delay_comp_mmf_zf} are finite because they are obtained after $10^4$ time slots, and it will increase to infinity with time.} To demonstrate this even clearer, in Fig.~\ref{fig:delay_increase} we show the evolution of the time-average delay with the number of time slots, showing that the delay obtained with the MMF and modified MMF grows linearly with time. 

\begin{figure}[t!]
	\centering
	\includegraphics[width=0.99\columnwidth]{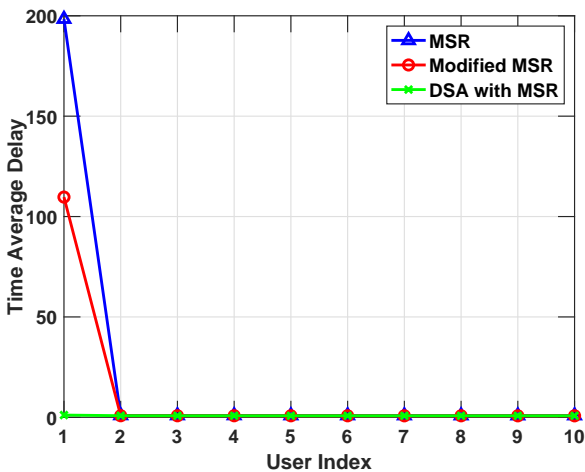}
	\caption{Time-average delay ($\times 10^{-5}\,$s/bit) of $K=10$ users with MRC  in co-located Massive MIMO. The results are obtained after $t=10^{4}$ time slots. All the users have the same data arrival rates, i.e., $B_{\max}=5\times\tau_c$, $p_k=0.5$ for all $k=1,\ldots, 10$.}
	\label{fig:delay_comp_msr}
\end{figure}

\begin{figure}[t!]
	\centering
	\includegraphics[width=0.99\columnwidth]{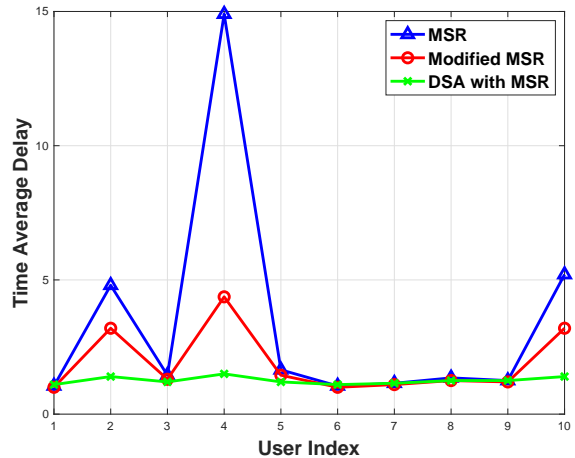}
	\caption{Time-average delay ($\times 10^{-5}\,$s/bit) of $K=10$ users with MRC  in co-located Massive MIMO, when user locations change over time. Same mobility model as in Fig.~\ref{fig:delay_comp_mmf_randomwalk}. Other parameters are the same as in Fig.~\ref{fig:delay_comp_msr}.}
	\label{fig:delay_comp_msr_diff_loc}
\end{figure}

In Figs.~\ref{fig:delay_comp_msr} and \ref{fig:delay_comp_msr_diff_loc}, we present the delay performance when the objective is to maximize the sum rate. The results are obtained with fixed user locations in Fig.~\ref{fig:delay_comp_msr}, and varying user locations in  Fig.~\ref{fig:delay_comp_msr_diff_loc}, respectively. 
In Fig.~\ref{fig:delay_comp_msr}, the arrival rates we choose are within the network capacity region, but user $1$ has higher data arrival rate than its optimal transmission rate derived with MSR in the infinite backlog case. If we do not take into account the bursty traffic, with conventional MSR scheme, user $1$ will have an unstable queue, which leads to infinite delay with time increasing. The DSA guarantees finite delay for all users, while the delay improvement comes at the price of sacrificing slightly the allocated rates for users with good channel conditions, which has very little impact on the delay of those users since their scheduled transmission rates are much higher than the data arrival rates. Similar to the MMF case, when the user are moving around, the DSA shows clear advantage in reducing the time-average delay compared to the baseline schemes.

\begin{figure}[t!]
	\centering
	\includegraphics[width=0.99\columnwidth]{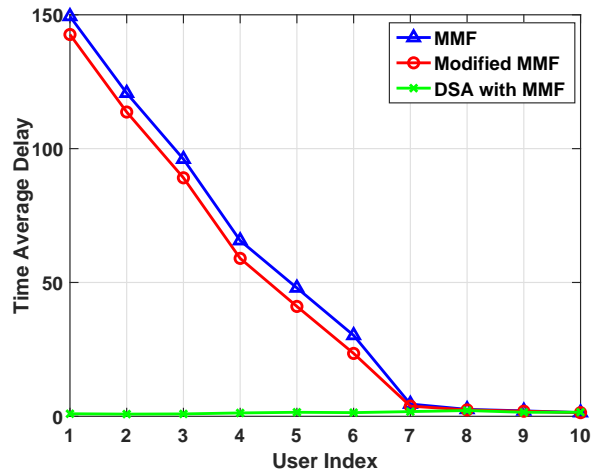}
	\caption{Time-average delay ($\times 10^{-5}\,$s/bit) of $K=10$ users in CF Massive MIMO after $t=10^{4}$ time slots. $B_{\max}=5\times\tau_c$, $p_1=0.65$, $p_{10}=0.5$, uniform spacing between $p_i$ and $p_{i+1}$ for $k=1,\ldots, 9$.}
	\label{fig:delay_cf}
\end{figure}

In Fig. \ref{fig:delay_cf}, we show the delay comparison in CF Massive MIMO, when the users have different data arrival rates. The arrival rate vector we choose falls inside the network capacity region, but for user $1-6$, their arrival rates exceed the optimal max-min rate obtained by the conventional MMF scheme. Therefore, both MMF and modified MMF schemes are unable to stabilize the network, and the users with high data arrival rates will have unbounded delay.

In summary, when all users have infinite data demand, the proposed dynamic scheduling and power control algorithm achieves the same long-term throughput as the conventional power control schemes. However, the cross-layer flow control in our algorithm can limit the data flow admitted to the network such that the transmission queues are stable. Furthermore, when the data arrival rates are within the network capacity region of the network, our algorithm can stabilize the system while achieving the optimal network fairness, thus guarantee finite delay. Conventional power control schemes which ignore the data traffic statistics can fail to stabilize the network even when it can be stabilized. As the result, some users will have infinite delay which increases with time.

\section{Conclusions}

In practice, the data arrivals in Massive MIMO systems will be random
and this fact makes the resource allocation problem rather different
from the infinite backlog scenario that has dominated the literature.
In this work, we studied cross-layer flow control and rate
allocation in uplink co-located and CF Massive MIMO systems. With the
help of Lyapunov optimization theory, we constructed a dynamic
scheduling algorithm that stabilizes the system and maximizes a
predefined utility function of the long-term user throughput. Compared to
the conventional deterministic power control schemes, our new
algorithm can substantially reduce the average delay experienced by
the users when their locations change over time, and guarantee finite
delay in scenarios where the conventional schemes fail to do that.

\appendix
\appendices

\subsection{Proof of Lemma \ref{lemma2}}
\label{appen1}
For the joint queues $\boldsymbol{\Theta}(t)=[\mathbf{Y}(t);\mathbf{Q}(t)]$, 
from \eqref{eq:lyapounov-function} and \eqref{eq:one-step-drift}, the one step Lyapunov drift is obtained as
\begin{align}
\Delta\big(\boldsymbol{\Theta}(t)\big)=~&\mathbb{E}[\mathcal{L}(\boldsymbol{\Theta}(t+1)-\mathcal{L}(\boldsymbol{\Theta}(t))|\boldsymbol{\Theta}(t)] \nonumber\\
=~&\mathbb{E}\Bigg\{\frac{1}{2}\sum\limits_{k=1}^{K}\big[Q_k^2(t+1)-Q_k^2(t)\big]\nonumber\\&+\frac{\eta}{2}\sum\limits_{k=1}^{K}\big[Y_k^2(t+1)-Y_k^2(t)\big]|\boldsymbol{\Theta}(t)\Bigg\}. \label{eq:drift2}
\end{align}
Recall that the queues update as follows.
\begin{align} 
Q_k(t+1)&=\max[Q_k(t)-R_k(t),0]+A_k(t), \nonumber \\
Y_k(t+1)&=\max[Y_k(t)-A_k(t),0]+\nu_{k}(t). \nonumber
\end{align}
We have
{\small
\begin{align}
&Q_k(t+1)^2\nonumber \\ \leq&~\left[Q_k(t)-R_k(t)\right]^2+A_k(t)^2+2A_k(t)\max[Q_k(t)-R_k(t),0]\nonumber \\
\leq &~~Q_k(t)^2+R_k(t)^2+A_k(t)^2-2Q_k(t)\big[R_k(t)-A_k(t)],
\end{align}}
and 
{\small
\begin{equation}
Y_k(t+1)^2\leq  Y_k(t)^2+A_k(t)^2+\nu_k(t)^2-2Y_k(t)\big[A_k(t)-\nu_k(t)].
\end{equation}}
From \eqref{eq:drift2}, we have
\begin{equation}
\begin{split}
\Delta\big(\boldsymbol{\Theta}(t)\big) \leq&~\frac{1}{2}\sum\limits_{k=1}^{K}\mathbb{E}\left[R_k(t)^2+A_k(t)^2|\boldsymbol{\Theta}(t)\right]\\&-\sum\limits_{k=1}^{K}\mathbb{E}[Q_k(t)\big(R_k(t)-A_k(t))|\boldsymbol{\Theta}(t)]\\
&+\frac{\eta}{2}\sum\limits_{k=1}^{K}\mathbb{E}\left[A_k(t)^2+\nu_k(t)^2|\boldsymbol{\Theta}(t)\right] \\
&-\eta\sum\limits_{k=1}^{K}\mathbb{E}[Y_k(t)\big(A_k(t)-\nu_k(t))|\boldsymbol{\Theta}(t)].
\end{split}
\label{eq:drift-bound1}
\end{equation}
From the system model and the constraints in \eqref{cons1} and \eqref{cons2}, we have $\mathbb{E}[A_k^2(t)|\boldsymbol{\Theta}(t)]\leq  A_{\max}^2$, $\mathbb{E}[\nu_k^2(t)|\boldsymbol{\Theta}(t)]\leq A_{\max}^2$, and $\mathbb{E}[R_k(t)^2]\leq R_{k,\max}^2$, where $R_{k,\max}$ is the maximum achievable rate of user $k$ (the rate that user $k$ has when it transmits with full power and all other users use zero power). Then we have
\begin{equation}
\begin{split}
&\Delta\left(\boldsymbol{\Theta}(t)\right)\\ &\leq~\frac{1}{2}\sum\limits_{k=1}^{K}R_{k,\max}^2+\frac{2\eta +1}{2}K A_{\max}^2-\sum\limits_{k=1}^{K}\mathbb{E}[Q_k(t)R_k(t)|\boldsymbol{\Theta}(t)]\\&-\!\sum\limits_{k=1}^{K}\!\mathbb{E}[A_k(t)\!\left(\eta Y_k(t)\!-\!Q_k(t)\right)\!|\boldsymbol{\Theta}(t)]\!+\!\eta\!\sum\limits_{k=1}^{K}\!\mathbb{E}[Y_k(t)\nu_k(t)|\boldsymbol{\Theta}(t)].
\end{split}
\end{equation}
Adding the penalty term $-V \mathbb{E}[f(\boldsymbol{\nu})|\boldsymbol{\Theta}(t)]$ to the one-step drift, the penalty-plus-drift is lowered bounded by
\begin{align}
&~\Delta\left(\boldsymbol{\Theta}(t)\right)\!-\! V \mathbb{E}[f(\boldsymbol{\nu})|\boldsymbol{\Theta}(t)]\nonumber \\
\leq & ~\frac{1}{2}\sum_{k=1}^{K}R_{k,\max}^2\!+\!\frac{2\eta+1}{2} K A_{\max}^2-\mathbb{E}\left[\sum_{k=1}^{K}Q_k(t) R_k(t)| \boldsymbol{\Theta}(t)\right]
\nonumber \\
&-\mathbb{E}\left[\sum_{k=1}^{K}A_k(t)\big(\eta Y_k(t)-Q_k(t)\big)| \boldsymbol{\Theta}(t)\right] \nonumber \\
&-\mathbb{E}\left[Vf(\boldsymbol{\nu})\!-\!\eta \sum\limits_{k=1}^{K}Y_k(t)\nu_k(t)| \boldsymbol{\Theta}(t)\right].
\label{eq:drift-penalty-bound}
\end{align}
Denote $C=\frac{1}{2}\sum\limits_{k=1}^{K}R_{k,\max}^2+\frac{2\eta+1}{2}K A_{\max}^2$, we obtain Lemma~\ref{lemma2}.

\end{document}